\documentclass[aps,twocolumn,prl]{revtex4}

\usepackage{epsfig}
\newcommand{\be}{\begin{equation}}
\newcommand{\ee}{\end{equation}}
\newcommand{\bea}{\begin{eqnarray}}
\newcommand{\eea}{\end{eqnarray}}
\newcommand{\ba}{\begin{array}}
\newcommand{\ea}{\end{array}}

\begin{document}
\title{Hierarchical Clustering Using Mutual Information}
\author{Alexander Kraskov, Harald St\"ogbauer, Ralph G. Andrzejak, and Peter Grassberger}
\affiliation{John-von-Neumann Institute for Computing, Forschungszentrum J\"ulich,
   D-52425 J\"ulich, Germany}

\date{\today}
\begin{abstract}
We present a method for hierarchical clustering of data called {\it mutual information clustering} (MIC)
algorithm. It uses mutual information (MI) as a similarity measure and exploits its grouping property: The MI
between three objects $X, Y,$ and $Z$ is equal to the sum of the MI between $X$ and $Y$, plus the MI between $Z$
and the combined object $(XY)$. We use this both in the Shannon (probabilistic) version of information theory
and in the Kolmogorov (algorithmic) version. We apply our method to the construction of phylogenetic trees from
mitochondrial DNA sequences and to the output of independent components analysis (ICA) as illustrated with the
ECG of a pregnant woman.
\end{abstract}

\maketitle

Classification or organizing of data is very important in all scientific disciplines and is fundamental for
understanding and learning \cite{jain-dubes}. Classification can be exclusive or overlapping, supervised or
unsupervised. In the following we will be interested only in exclusive unsupervised classification, called
clustering.

An instance of a clustering problem consist of a set of objects and a set of properties (called characteristic
vector) for each object. The goal of clustering is separation of objects into groups using only the
characteristic vectors. Cluster analysis organizes data either as a single grouping of individuals into
non-overlapping clusters or as a hierarchy of nested partitions. The latter is called hierarchical clustering
(HC). Because of wide spread of applications, there are a large variety of different clustering methods in
usage, see e.g. \cite{jain-dubes} for an overview.

 The crucial point of all clustering algorithms is the choice
of a {\it proximity measure}. This is obtained from the characteristic vectors and can be either an indicator
for similarity or dissimilarity. In the latter case it is convenient but not obligatory to satisfy the standard
axioms of a metric (positivity, symmetry, and triangle inequality). Among HC methods one should distinguish
between those where one uses the characteristic vectors only at the first level of the hierarchy and derives the
proximities between clusters from the proximities of their constituents, and methods where the proximities are
calculated each time from their characteristic vectors. The latter strategy (which is used also in the present
paper) allows of course for more flexibility but might also be computationally more costly.

Quite generally, the ``objects" to be clustered can be either single (finite) patterns (e.g. DNA sequences) or
random variables, i.e. {\it probability distributions}. In the latter case the data are usually supplied in form
of a statistical sample, and one of the simplest and most widely used similarity measures is the linear
(Pearson) correlation coefficient. But this is not sensitive to nonlinear dependencies which do not manifest
themselves in the covariance and can thus miss important features. This is in contrast to mutual information
(MI) which is also singled out by its information theoretic background \cite{cover-thomas}. Indeed, MI is zero
only if the two random variables are strictly independent.

Another important feature of MI is that it has also an ``algorithmic" cousin, defined within algorithmic
(Kolmogorov) information theory \cite{li-vi} which measures the similarity between individual objects. For a
thorough discussion of distance measures based on algorithmic MI and for their application to clustering, see
\cite{li1,li2}.

Essential for the present application is the {\it grouping property} of MI,
\be
   I(X,Y,Z) = I(X,Y) + I((X,Y),Z).                        \label{group}
\ee
Within Shannon information theory this is an exact theorem, while it is true in the algorithmic version up to
the usual logarithmic correction terms \cite{li-vi}. Since $X,Y,$ and $Z$ can be themselves composite,
Eq.(\ref{group}) can be used recursively for a cluster decomposition of MI. This motivates the main idea of our
clustering method: instead of using e.g. centers of masses in order to treat clusters like individual objects in
an approximative way only, we treat them exactly like individual objects when using MI as proximity measure and
We thus propose the following scheme for clustering $n$ objects with MIC:\\
(1) Compute a proximity matrix based on pairwise mutual informations; assign $n$ clusters
such that each cluster contains exactly one object;\\
(2) find the two closest clusters $i$ and $j$; \\
(3) create a new cluster $(ij)$ by combining $i$ and $j$; \\
(4) delete the lines/columns with indices $i$ and $j$ from the proximity matrix, and add one line/column
containing the proximities between cluster $(ij)$ and all
other clusters; \\
(5) if the number of clusters is still $>2$, goto (2); else join the two clusters and stop.

{\bf Shannon Theory:} Here, $X\equiv X_1, Y\equiv X_2, \ldots$ are random variables. If they are discrete,
entropies are defined as usual $H(X) = - \sum_ip_i(X) \log p_i(X)$ etc. The MI is defined as
\be
   I(X_1,\ldots, X_n) = \sum_{k=1}^n H(X_k) - H(X_1,\ldots, X_n).
\ee
Eq.(\ref{group}) can be checked easily, together with its generalization to arbitrary groupings. It means that
MI can be {\it decomposed into hierarchical levels}. By iterating it, one can decompose $I(X_1\ldots X_n)$ for
any $n>2$ and for any partitioning of the set $(X_1\ldots X_n)$ into the MIs between elements within one cluster
and MIs between clusters.

For continuous variables with densities $\mu_X$ etc., one first introduces some binning (`coarse-graining'), and
applies the above to the binned variables. If $x$ is a vector with dimension $m$ and each bin has Lebesgue
measure $\Delta$, then $p_i(X) \approx \mu_X(x)\Delta^m$ with $x$ chosen suitably in bin $i$, and
\be
   H_{\rm bin}(X) \approx \tilde{H}(X) - m \log \Delta
\ee
where the {\it differential entropy} is given by
\be
   \tilde{H}(X) = -\int dx \;\mu_X(x) \log \mu_X(x).
\ee
Notice that $H_{\rm bin}(X)$ is a true (average) information and is thus non-negative, but $\tilde{H}(X)$ is not
an information, can be negative, and is not invariant under homeomorphisms $x\to \phi(x)$.

Joint entropies, conditional entropies, and MI are defined as above, with sums replaced by integrals. Like
$\tilde{H}(X)$, joint and conditional entropies are neither positive (semi-)definite nor invariant. But MI,
defined as
\be
   I(X,Y) = \int\!\!\!\int dx dy \;\mu_{XY}(x,y) \;\log{\mu_{XY}(x,y)\over \mu_X(x)\mu_Y(y)}\;,
   \label{mi}
\ee
is non-negative and invariant under $x\to \phi(x)$ and $y\to \psi(y)$. It is (the limit of) a true information,
\be
   I(X,Y) = \lim_{\Delta\to 0} [H_{\rm bin}(X)+H_{\rm bin}(Y)-H_{\rm bin}(X,Y)].
\ee

In applications, one usually has the data available in form of $N$ sample points $(x_i,y_i), \, i=1,\ldots N$
which are assumed to be i.i.d. realizations. There exist numerous algorithms to estimate $I(X,Y)$ and entropies.
We use in the following the MI estimators proposed recently in Ref.~\cite{mi}, and we refer to this paper for a
review of alternative methods.

{\bf Algorithmic Information Theory:} In contrast to Shannon theory where the basic objects are random variables
and entropies are {\it average} informations, algorithmic information theory deals with individual symbol
strings and with the actual information needed to specify them. To ``specify" a sequence $X$ means here to give
the necessary input to a universal computer $U$, such that $U$ prints $X$ on its output and stops. The analogon
to entropy, called here usually the {\it complexity} $K(X)$ of $X$, is the minimal length of an input which
leads to the output $X$, for fixed $U$. It depends on $U$, but it can be shown that this dependence is weak and
can be neglected in the limit when $K(X)$ is large \cite{li-vi}.

Let us denote the concatenation of two strings $X$ and $Y$ as $XY$. Its complexity is $K(XY)$. It is intuitively
clear that $K(XY)$ should be larger than $K(X)$ but cannot be larger than the sum $K(X)+K(Y)$. Finally, one
expects that $K(X|Y)$, defined as the minimal length of a program printing $X$ when $Y$ is furnished as
auxiliary input, is related to $K(XY)-K(Y)$. Indeed, one can show \cite{li-vi} (again within correction terms
which become irrelevant asymptotically) that
\be
   0 \leq K(X|Y) \simeq K(XY)-K(Y) \leq K(X).
\ee
Notice the close similarity with Shannon entropy. The algorithmic information in $Y$ about $X$ is finally
\be
   I_{\rm alg}(X,Y) = K(X) - K(X|Y) \simeq K(X)+K(Y)-K(XY),
\ee
and similarly for more than two strings. Within the same additive correction terms, one shows that it is
symmetric, $I_{\rm alg}(X,Y) =I_{\rm alg}(Y,X)$, and can thus serve as an analogon to mutual information.

$K(X)$ is in general not computable. But one can easily give upper bounds: The length of any input which
produces $X$ (e.g. by spelling it out verbatim) is an upper bound. Improved upper bounds are provided by any
file compression algorithm.

{\bf MI-Based Distance Measures:} When comparing objects with different marginal or joint informations, it seems
intuitively clear that one should prefer {\it relative} distances over absolute ones, in order to minimize the
dependence on the total information. We here use
the quantity \cite{li1,cluster}
\be
   D(X,Y) = 1 - \frac{I(X,Y)}{H(X,Y)}                  \label{eq:dist}
\ee
which is a metric , with $D(X,X)=0$ and $D(X,Y)\leq 1$ for all pairs $(X,Y)$. The algorithmic version is also
{\it universal}: If $X\approx Y$ according to any non-trivial distance measure, then $X\approx Y$ also according
to $D$.

A difficulty appears in the Shannon framework, if we deal with continuous random variables. As we mentioned
above, $\tilde{H}(X,Y)$ is not invariant under homeomorphisms (including rescalings) and not even positive
definite, while $H_{\rm bin}$ diverges when $\Delta\to 0$. We thus modified Eq.(\ref{eq:dist}) by replacing
$H(X,Y)$ by $H_{\rm bin}(X,Y)$ and replacing $D(X,Y)$ by the similarity measure
\be
   S(X,Y) = \lim_{\Delta\to 0} (D(X,Y)-1)\log\Delta  = \frac{I(X,Y)}{m_x+m_y}.
                                         \label{S}
\ee

{\bf A Phylogenetic Tree for Mammals:} We study the mitochondrial DNA of a group of 34 mammals (see Fig.~1). The
same data \cite{Genebank} had previously been analyzed in \cite{li1,Reyes00}. This group includes among others
some rodents, ferungulates, and primates.

Obviously we are here dealing with the algorithmic version of information theory, and informations are estimated
by lossless data compression. For constructing the proximity matrix between individual taxa, we proceed
essentially as in Ref.~\cite{li1}, using the special compression program GenCompress \cite{GenComp}.

\begin{figure}
  \begin{center}
    \psfig{file=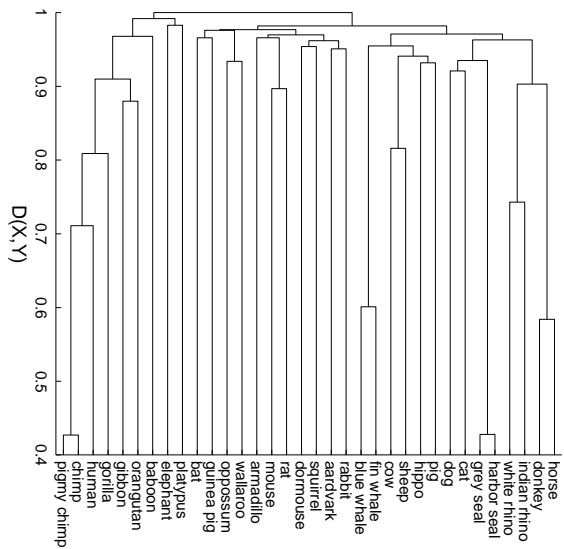,height=75mm,angle=270}
    \caption{Phylogenetic tree for 34 mammals.
       The heights of nodes are the distances
       between the joining daughter clusters.}
    \label{phylotree}
  \end{center}
  \vspace{-8mm}
\end{figure}

In Ref.~\cite{li1}, this proximity matrix was then used as the input to a standard HC algorithm
(neighbour-joining and hypercleaning) to produce an evolutionary tree. Instead we use the MIC algorithm with
distance $D(X,Y)$. The joining of two clusters is obtained by simply concatenating the DNA sequences. There is
of course an arbitrariness in the order of concatenation sequences: $XY$ and $YX$ give in general compressed
sequences of different lengths. But we found this to have negligible effect on the evolutionary tree.

The overall structure of this tree closely resembles the one shown in Ref.~\cite{Reyes00}. All primates are
correctly clustered and also the relative order of the ferungulates (blue whale to horse) is in accordance with
Ref.~\cite{Reyes00}. On the other hand, there are a number of connections which obviously do not reflect the
true evolutionary tree, see for example the guinea pig with bat and elephant with platypus.
%
%
But the latter two, inspite of being joined together, have a very large distance from each other, thus their
clustering just reflects the fact that neither the platypus nor the elephant have other close relatives in the
sample. All in all, however, already the results shown in Fig.~1 capture surprisingly well the overall structure
shown in Ref. \cite{Reyes00}. Dividing MI by the total information is essential for this success. If we had used
the non-normalized $I_{\rm alg}(X,Y)$ itself, the clustering algorithm used in \cite{li1} would not change much,
since all 34 DNA sequences have roughly the same length. But our MIC algorithm would be completely screwed up:
After the first cluster formation, we have DNA sequences of very different lengths, and longer sequences tend
also to have larger MI, even if they are not closely related.

The concatenation of $X$ and $Y$ will of course not lead to a plausible sequence of the common ancestor, but it
{\it optimally represents the information} about it. This information is essential to find the correct way
through higher hierarchy levels of the evolutionary tree, and it is preserved in concatenating.

{\bf Clustering of Minimally Dependent Components in an Electrocardiogram:} As our second application we choose
a case where Shannon theory is the proper setting. We show in Fig.~2 an ECG recorded from the abdomen and thorax
of a pregnant woman \cite{ECGdata}. It is already seen from this graph that there are at least two important
components in this ECG: the heartbeat of the mother and of the fetus. In addition there is noise from various
sources (muscle activity, measurement noise, etc.). While it is easy to detect anomalies in the mother's ECG
from such a recording, it would be difficult to detect them in the fetal ECG.

As a first approximation we can assume that the total ECG is a linear superposition of several independent
sources (mother, child, noise$_1$, noise$_2$,...). A standard method to disentangle such superpositions is {\it
independent component analysis} (ICA) \cite{ICA}. There, one tries to recover the sources by means of linear
transformation $s_i(t) = \sum_{j=1}^n W_{ij} x_j(t)$, where $W_{ij}$ is determined by minimizing the estimated
MI between the $s_i$.

In reality things are not so simple. For instance, the sources might not be independent, the number of sources
(including noise sources!) might be different from the number of channels, and the mixing might involve delays.
For the present case this implies that the heartbeat of the mother is seen in several reconstructed components
$s_i$, and that the ``independent" components are not independent at all. In particular, all components $s_i$
which have large contributions from the mother form a cluster with large intra-cluster MIs and small
inter-cluster MIs. The same is true for the fetal ECG, albeit less pronounced. To obtain clean recordings of the
fetal and maternal ECGs, we proceeded as follows \cite{Harald}.

Since we expect different delays in the different channels, we first used Takens delay embedding \cite{Takens80}
with time delay 0.002$\,$s and embedding dimension 3, resulting in $24$ channels. We then formed 24 linear
combinations $s_i(t)$. We use the MI estimator \cite{mi}, for details see \cite{cluster}. Five of the resulting
least dependent components contain strong contributions of the mother's heartbeat, three are dominated by the
fetus. The rest contains mostly noise \cite{cluster}.

In plotting the actual dendrogram (Fig.~\ref{ClustECG}) we used $S(X,Y)$ for the cluster analysis but used the
MI of the clusters to determine the height at which the two branches join. The MI of the first five channels,
e.g., is $\approx 1.44$~nats, while that of channels 6 to 8 is $\approx 0.3$~nats. For any two clusters (tuples)
$X=X_1\ldots X_n$ and $Y=Y_1\ldots Y_m$ one has $I(X,Y) \geq I(X)+I(Y)$. This guarantees, if the MI is estimated
correctly, that the tree is drawn properly. The two slight glitches (when clusters (1 - 14) and (15 - 18) join,
and when (21 - 22) is joined with 23) result from small errors in estimating MI. They do in no way effect our
conclusions.

\begin{figure}
  \begin{center}
    \psfig{file=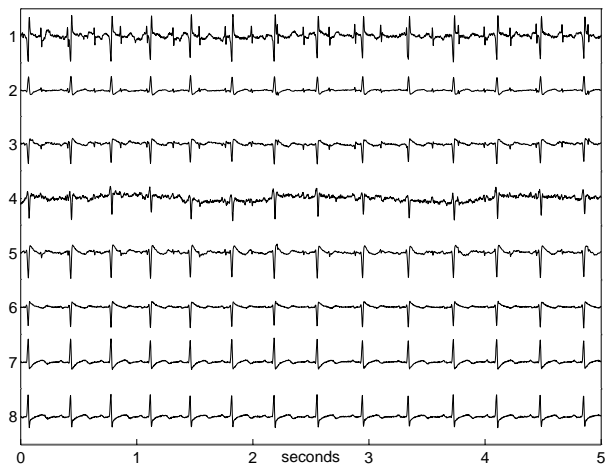,width=8cm}
    \caption{ECG of a pregnant woman (sampling rate 500 Hz).}
    \label{ICAECG0}
  \end{center}
\end{figure}

\begin{figure}
  \begin{center}
    \psfig{file=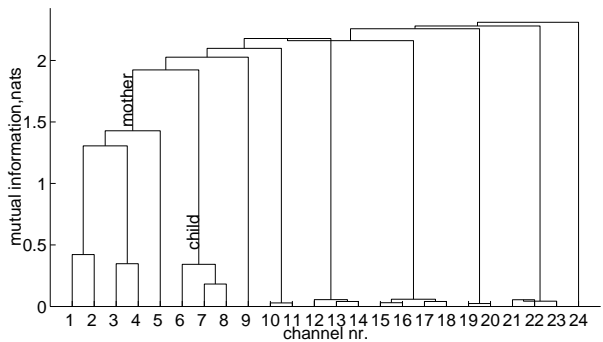,width=8cm,angle=0}
    \caption{Dendrogram for least dependent components.}
    \label{ClustECG}
  \end{center}
  \vspace{-8mm}
\end{figure}

\begin{figure}[t]
  \begin{center}
    \psfig{file=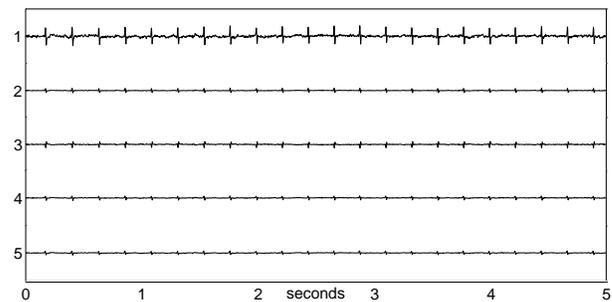,width=8cm}
    \caption{ECG where all contributions except those of the child cluster have
     been removed.}
    \label{reconstruct}
  \end{center}
  \vspace{-8mm}
\end{figure}

In Fig.~\ref{ClustECG} one can clearly see two big clusters corresponding to the mother and to the child. There
are also some small clusters which should be considered as noise. For reconstructing the mother and child
contributions to Fig.~\ref{ICAECG0}, we have to decide on one specific clustering from the entire hierarchy. We
decided to make the cut at inter-cluster MI equal to 0.1, i.e. two clusters $X$ and $Y$ are joined whenever
$I((X),(Y)) \equiv I(X,Y)-I(X)-I(Y) \geq 0.1$. Reconstructing the first five traces of the original ECG from the
child components only, we obtain Fig.~\ref{reconstruct}.

In summary, we have shown that MI can not only be used as a proximity measure in clustering, but that it also
suggests a conceptually very simple and natural hierarchical clustering algorithm. We do not claim that this
algorithm, called {\it mutual information clustering} (MIC), is always superior to other algorithms. Indeed, MI
is in general not easy to estimate. Obviously, when only crude estimates are possible, also MIC will not give
very good results. But as MI estimates are becoming better, also the results of MIC should improve. The present
paper was partly triggered by the development of a new class of MI estimators for continuous random variables
which have very small bias and also rather small variances \cite{mi}.

We have illustrated our method with two applications, one from genetics and one from cardiology. For neither
application MIC might give optimal clustering, but it seems promising that one common method gives decent
results for both, although they are very different.

The results of MIC should improve, if more data become available. This is trivial, if we mean by that longer
time sequences in the application to ECG, and longer parts of the genome. It is less trivial that we expect MIC
to make fewer mistakes in a phylogenetic tree, when more species are included. The reason is that close-by
species will be correctly joined anyhow, and families -- which now are represented only by single species and
thus are poorly characterized -- will be much better described by the concatenated genomes if more species are
included.

We would like to thank Arndt von Haesseler, Walter Nadler and Volker Roth for many useful discussions.

\end{document}